# Superconducting Nanowire Single-Photon Detector (SNSPD) with 3D-Printed Free-Form Microlenses

Yilin Xu,[1,3,*] Artem Kuzmin,[2] Emanuel Knehr,[2] Matthias Blaicher,[1,3] Konstantin Ilin,[2] Philipp-Immanuel Dietrich,[1,3,4] Wolfgang Freude,[1] Michael Siegel,[2] and Christian Koos[1,3,4,*]

[1]*Institute of Photonics and Quantum Electronics (IPQ), Karlsruhe Institute of Technology (KIT), Engesserstrasse 5, 76131 Karlsruhe, Germany*
[2]*Institute of Micro- and Nanoelectronic Systems (IMS), KIT, Hertzstrasse 16, 76187 Karlsruhe, Germany*
[3]*Institute of Microstructure Technology (IMT), KIT, Hermann-von-Helmholtz-Platz 1, 76344 Eggenstein-Leopoldshafen, Germany*
[4]*Vanguard Automation GmbH, Gablonzer Strasse 10, 76185 Karlsruhe, Germany*

*\*yilin.xu@kit.edu; christian.koos@kit.edu*

**Abstract:** We present an approach to increase the effective light-receiving area of superconducting nanowire single-photon detectors (SNSPD) by free-form microlenses. These lenses are printed *in situ* on top of the sensitive detector areas using high-resolution multi-photon lithography. We demonstrate a detector based on niobium-nitride (NbN) nanowires with a 4.5 µm × 4.5 µm sensitive area, supplemented with a lens of 60 µm diameter. For a plane-wave-like free-space illumination at a wavelength of 1550 nm, the lensed sensor has a 100-fold increased effective collection area, which leads to a strongly enhanced system detection efficiency without the need for long nanowires. Our approach can be readily applied to a wide range of sensor types. It effectively overcomes the inherent design conflict between high count rate, high timing accuracy, and high fabrication yield on the one hand and high collection efficiency through a large effective detection area on the other hand.

## 1. Introduction

Superconducting nanowire detectors [1] are key to many applications that require single-photon detection in the optical and near-infrared spectral region. Superconducting nanowire single-photon detectors (SNSPD) are fabricated from a thin superconducting film patterned to a stripe (nanowire), which is biased close to the critical current where superconductivity disappears. If any extra energy, e.g., from a photon, is absorbed by the nanowire, a so-called hot spot appears, i.e., a region with suppressed superconductivity. The nanowire then switches to the normal conducting state, and a voltage pulse from this event can be detected in an external circuit [2]. Despite the requirement of cryogenic operating temperatures, SNSPD are attractive due to their ability to cover a broad spectral range from ultra-violet (UV) to mid-infrared with a quantum efficiency of up to 98 % [3]. Picosecond timing jitter [4], gigahertz photon count rates (PCR) [5], and sub-1 Hz dark count rates (DCR) are further advantages. Promising results with SNSPD were already obtained in laser ranging (LiDAR) [6,7], spectroscopy [8–10], quantum key distribution [11,12], as well as in particle and nuclear physics [13]. Further application fields are deep-space communications [14] and integrated quantum photonics [15].

In most cases, SNSPD consist of meander-like nanowires with typical widths of the order of 100 nm that are fabricated on a plane substrate and illuminated from a direction normal to the substrate plane to avoid technically complex and lossy coupling of photons into integrated optical waveguides. This leads to design conflicts regarding the nanowire length: While high PCR, low DCR, and low timing jitter require a short nanowire, the system detection efficiency (SDE) crucially depends on the covered area and thus calls for a long nanowire. In addition, large-area SNSPD with long nanowires are prone to random fabrication defects, thereby reducing the process yield. SNSPD based on niobium nitride (NbN) are widely used due to rather high operating temperatures up to 5 K, and have been demonstrated with active areas of, e.g., 26 µm  290 µm = 7540 µm² and area fill factors of up to 0.28 [16]. In these devices, however, the PCR is typically limited to less than 10 MHz due to the high kinetic inductance of the underlying 20-mm-long nanowire. In addition, the timing accuracy of such devices degrades with increasing detector length due to the so-called geometrical jitter [17], a random delay of an electrical pulse propagating from different absorption sites along the nanowire. On the other hand, maximum PCR of 2 GHz have been shown in SNSPD with 500 µm-long nanowires, but the active area of these devices is usually less than 100 µm² [18], which leads to rather low SDE in typical applications.

In this paper we show that this design conflict can be overcome by exploiting advanced 3D laser lithography for *in-situ* fabrication of large-area light-collection lenses on top of compact SNSPD with short nanowires. In our proof-of-concept experiments, we show 3D-printed freeform lenses on top of high-PCR SNSPD made from a 100 nm-wide NbN stripe. The 3D-printed lenses focus the incident light to the associated SNSPD with a lateral precision better than 100 nm and offer effective collection areas of more than 2000 µm², while keeping the nanowire length as short as 100 µm. This leads to short reset times of less than 2 ns, thereby enabling peak PCR of hundreds of MHz, which might be further increased to a few GHz. Our approach is general and can be transferred to extended SNSPD arrays that combine high detection efficiency with high peak PCR and high fabrication yield.

## 2. Improving detection efficiency of SNSPD by 3D-printed microlenses

The concept of 3D-printed microlenses on top of an SNSPD is illustrated in Fig. 1. Figure 1a shows a schematic view of a 16-pixel SNSPD array with hexagonal arrangement. The SNSPD array is combined with an associated array of 3D-printed microlenses, each of which collects incoming light from an effective collection area $A_C$ and focuses it to a spot within the active area $A_D$ of the corresponding SNSPD. The inset of Fig. 1a shows a magnified view of a single SNSPD, which, in our case, has an active detection area of $A_D = 4.5 \times 4.5\,\mu\text{m}^2$ into which a circle with radius $r_D$ can be inscribed. Figure 1b shows a schematic cross-section through an individual lens with apex height $h_0$ having a rotationally symmetric lens surface that is described in cylindrical coordinates by the function $h(r)$. The lens geometry is characterized by a physical aperture with radius $r_A$, which denotes the distance from the optical axis at which the lens profile is clipped. Note that the radius $r_C$ of the effective collection area $A_C = \pi r_C^2$ may be additionally limited by strong Fresnel reflection at the increasingly steep lens surface in the regions that are further away from the optical axis, i.e., $r_C \leq r_A$. The effective collection area of a certain lens is defined as the area oriented perpendicular to the direction of an incoming plane wave, which would collect the same optical power as the lens itself, see Appendix A.1 for details.

To quantify the performance of our 3D-printed lenses, we first introduce a quantitative description of the system detection efficiency (SDE), which may be improved by the increased collection area $A_C$ of the 3D-printed microlenses. In the following, $R_i$ denotes the rate of photons which are incident on an input aperture of the detection system. This input aperture may be defined by the end-face of an optical fiber or, in case of free-space illumination, by an optical window in the cryostat. The SDE is defined as the ratio of the average photon count rate PCR captured by the SNSPD and the photon rate $R_i$ incident on the input aperture, $\text{SDE} = \text{PCR}/R_i$. We further introduce the rate $R_r$ of photons received by the active detector area $A_D$ and the rate $R_a$ of photons absorbed by the detector. The system detection efficiency can then be represented as a product of the optical coupling efficiency $\text{OCE} = R_r/R_i$, the absorption efficiency $\text{ABS} = R_a/R_r$, and the intrinsic detection efficiency $\text{IDE} = \text{PCR}/R_a$, where IDE represents the fraction of absorbed photons that lead to hot spots and therefore cause observable detector pulses. The system detection efficiency can thus be written as

$$\text{SDE} = \text{OCE} \times \text{ABS} \times \text{IDE}, \quad \text{SDE} = \frac{\text{PCR}}{R_i}, \quad \text{OCE} = \frac{R_r}{R_i}, \quad \text{ABS} = \frac{R_a}{R_r}, \quad \text{IDE} = \frac{\text{PCR}}{R_a}, \quad (1)$$

where all the efficiencies SDE, OCE, ABS, and IDE depend on the photon energy.

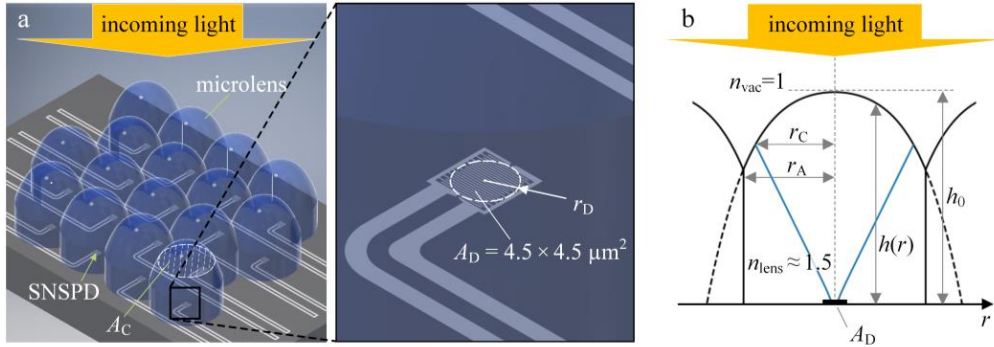

**Fig. 1.** Concept of 3D-printed microlenses on top of SNSPD. **(a)** Schematic of a 16-pixel SNSPD array with a corresponding microlens array in hexagonal arrangement. Each microlens collects incoming light from an effective collection area $A_C$ (hatched area) and focuses it to a spot within the active area $A_D$ of the respective SNSPD, into which a circle with radius $r_D$ can be inscribed. **(b)** Schematic cross-section through an individual lens with apex height $h_0$. Plane-wave-like light incident along the surface normal of the SNSPD substrate is focused to a spot with a radius smaller than $r_D$, where $r_D$ describes the radial extension of the SNSPD. The rotationally-symmetric lens surface is described in cylindrical coordinates by the function $h(r)$, where $r$ denotes the radial coordinate. The lens surface is clipped at the radius $r_A$, thereby defining the lens aperture. The radius $r_C$ of the effective collection area $A_C = \pi r_C^2$ may be additionally limited by strong Fresnel reflection at the increasingly steep lens surface in the regions that are further away from the optical axis, i.e., $r_C \leq r_A$, see Appendix A.1 for a quantitative description.

For a plane-wave-like illumination, the microlenses improve the system detection efficiency SDE foremost by effectively enlarging the detector area from $A_D$ to $A_C$. Photons are thus extracted from a larger portion of the input aperture, i.e., the optical coupling efficiency OCE is improved. In the following, we compare a single lensed detector (subscript "lens") to an identical reference detector without lens (subscript "ref"). The improvement of the optical coupling efficiency OCE is then described by the effective lens gain $G_D = \text{OCR}_{\text{lens}}/\text{OCR}_{\text{ref}} = A_C/A_D$. Assuming further that both detectors have equal absorption and intrinsic detection efficiencies, $\text{ABS}_{\text{lens}} = \text{ABS}_{\text{ref}}$ and $\text{IDE}_{\text{lens}} = \text{IDE}_{\text{ref}}$, we find according to Eq. (1) that $G_D$ also describes the increase of PCR and the improvement in SDE,

$$\frac{\text{SDE}_{\text{lens}}}{\text{SDE}_{\text{ref}}} = \frac{\text{PCR}_{\text{lens}}/R_i}{\text{PCR}_{\text{ref}}/R_i} = \frac{\text{OCE}_{\text{lens}}}{\text{OCE}_{\text{ref}}} = G_D, \qquad G_D = \frac{A_C}{A_D} = \frac{\pi r_C^2}{A_D}. \quad (2)$$

Note that fabrication defects and thermal fluctuations deteriorate the IDE in SNSPD with long nanowires [19]. Hence, the SDE of a lensed detector with collection area $A_C$ should be higher than the one of a long-nanowire SNSPD that covers the same equivalent area – in addition to the improved timing accuracy and detector speed [18]. Note also that, in practical applications, the propagation direction of the incident light might be misaligned with respect the optical axis of the 3D-printed lens by an angle $\gamma$ and that the improvement of the OCE by microlenses is subject to a fundamental tradeoff between the collection area $A_C$ and the maximum tolerable angular misalignment $\gamma_{\text{max}}$. This aspect is discussed in more detail in the following section.

## 3. Design of 3D-printed microlenses

As a first step of the design procedure, we consider a plane-wave illumination and use a simple ray-optics model to design a lens surface that focusses the incoming light to a single point in the center of the SNSPD. Based on this lens design, we then use a wave-optics model to estimate the achievable spot size on the SNSPD, and we derive analytical expressions to quantify the dependence of the effective lens gain $G_D$ and the maximum tolerable angular misalignment $\gamma_{max}$ of the illumination on the lens size. These considerations are followed by more detailed numerical simulations.

### 3.1 Analytic considerations

For the ray-optics lens design, we consider incident rays parallel to the optical axis of a rotationally symmetric lens, which is surrounded by vacuum (refractive index $n_{vac} = 1$), see Fig. 2a. We use cylindrical coordinates to describe the lens shape by the dependence of the height $h(r)$ on the radial coordinate $r$. We consider a ray impinging on the lens surface at an angle $\alpha_{vac}$ with respect to the local lens surface normal, see Fig. 2a, and we denote the corresponding angle inside the lens with $\alpha_{lens}$, which is connected to $\alpha_{vac}$ by Snell's law, $n_{lens} \sin \alpha_{lens} = n_{vac} \sin \alpha_{vac}$. We can then express the propagation angle $\theta = \alpha_{vac} - \alpha_{lens}$ of the internal ray with respect to the optical axis,

$$\theta = \alpha_{vac} - \arcsin\left(\frac{n_{vac}}{n_{lens}} \sin(\alpha_{vac})\right) < \theta_{max}, \quad (3)$$

where the maximum ray angle $\theta_{max}$ inside the bulk of the lens is limited by the refractive index of the lens, since $\alpha_{vac}$ cannot exceed $\pi/2$,

$$\theta_{max} = \frac{\pi}{2} - \arcsin\left(\frac{n_{vac}}{n_{lens}}\right). \quad (4)$$

Note that the maximum ray angle corresponds to the case of total internal reflection at the lens surface when considering a ray with reversed propagation direction from the inside of the lens to the outside. The optimum lens shape within this approximation of geometric optics is a spheroid, i.e., an ellipsoid that is rotationally symmetric with respect to the optical axis [20,21]. The two foci $S_1$ and $S_2$ of the spheroid are stacked vertically on the optical axis with the SNSPD placed in the lower one, see Fig. 2a. The ratio $\xi$ between the major (vertical) half-axis $b$ and the minor (horizontal) half-axis $a$ of the spheroid is related to the refractive index of the lens [20],

$$\xi = \frac{b}{a} = \frac{n_{lens}}{\sqrt{n_{lens}^2 - n_{vac}^2}}. \quad (5)$$

The distance $d$ between the center of the spheroid and any of the two foci is also referred to as the linear eccentricity

$$d = eb \quad (6)$$

which depends on the eccentricity

$$e = \sqrt{1 - (a/b)^2} = \sqrt{1 - 1/\xi^2}. \quad (7)$$

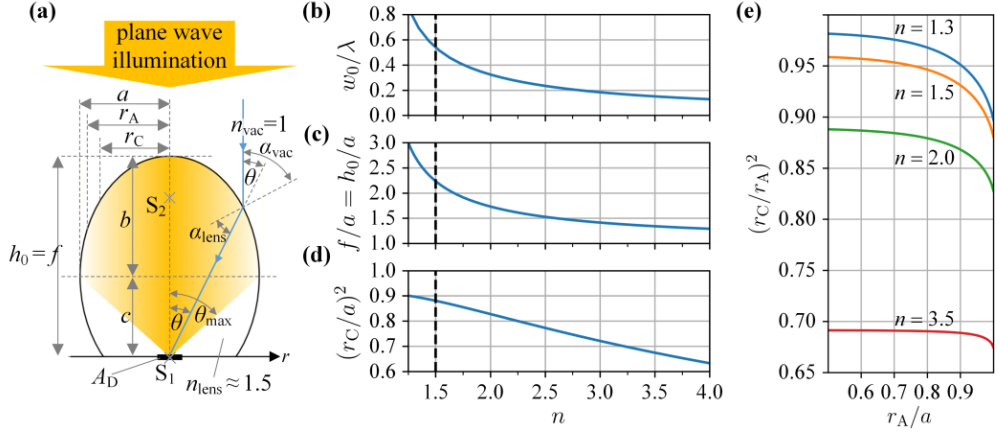

**Fig. 2.** Design procedure for a lensed SNSPD. **(a)** Cross section of the idealized spheroidal lens shape, which can be derived from ray-optical considerations [20,21]. The two foci $S_1$ and $S_2$ of the spheroid are stacked vertically. The SNSPD with area $A_D$ is located in the lower focus $S_1$, and the ratio $\xi = b/a$ between the half-axes $b$ and $a$ is fixed by the refractive index of the lens material, see Eq. (5). The apex height equals the material-sided focal distance $f$ and is given by Eq. (8). An exemplary ray (blue) impinges on the lens surface at a radial position $r$ with an angle $\alpha_{vac}(r)$ against the local surface normal and is refracted to an angle $\alpha_{lens}(r)$ within the lens. The associated angle to the optical axis is denoted as $\theta(r)$. The maximum possible aperture angle $\theta_{max}$ corresponds to the case of total internal reflection at the lens surface when considering a ray with reversed propagation direction from the inside of the lens to the outside, see Eq. (3). Both the effective collection radius $r_C$, see Appendix A.1 for details, and the focal distance $f$ are proportional to $a$. The size of the lens should thus be chosen as large as the required angular alignment tolerance permits, see Eq. (9). **(b)** Achievable second-moment-radius $w_0 = 2\sigma$ of a spot with an approximately Gaussian intensity distribution vs. refractive index $n$ of the lens material, see Appendix A.2 for details. The vertical dashed line indicates the typical available refractive index of $n_{lens} \approx 1.5$ for 3D-printed polymer lenses. **(c)** Ratio of focal length $f$ and minor half axis $a$ vs. refractive index $n$. For highest possible lens gain, Eq. (2), the half-axis $a$ should be chosen as large as possible while still respecting the upper limit for the focal distance $f$, which can be estimated through Eq. (9) based on the known spot size $w_0$, see Subfigure (b), the known detector size $r_D$, and the desired angular alignment tolerance. **(d)** Ratio of the effective collection are $\pi r_C^2$ and the geometrical cross-section $\pi a^2$ of the lens vs. refractive index $n$. The ratio decreases with increasing refractive index due to strong Fresnel reflections at the lens surface. **(e)** Effect of clipping the lenses, e.g., when integrated into a two-dimensional lens pattern. For simplicity, we assume that the pitch of the detectors can essentially be chosen freely and that the lenses are clipped circularly at an aperture radius $r_A < a$. The clipping removes the outer strongly inclined regions of the lens surfaces, which are subject to higher Fresnel reflections, such that strong clipping leads to an increased effective collection efficiency of the overall lens array. We find that the value of $(r_C/r_A)^2$ does not increase significantly with decreasing $r_A$ as soon as $r_A < 0.7 \times a$, i.e., clipping the lens surfaces to even smaller aperture radii does not pay out any more. This estimate helps to determine the number of detectors that are needed to realize a high-fill-factor array.

The material-sided focal distance $f$ of the spheroid lens equals the apex height $h_0$, see Fig. 2a, and scales with the size of the half axes as

$$f = h_0 = b + d = (1+e) \times b = \xi(1+e) \times a. \quad (8)$$

For such an spheroidal lens, we thus find that both the effective collection radius $r_C$, see Appendix A.1 for a details, and the focal distance $f$ are proportional to the horizontal half-axis $a$. By linear scaling of the lens, an arbitrarily high collection radius $r_C$ and thus an arbitrarily high lens gain $G_D$ can be achieved, see Eq. (2), provided that light is incident only along the major axis of the spheroid. In many practical applications, however, propagation direction of the incident light might be misaligned with respect the optical axis of the 3D-printed lens by an angle $\gamma$, which translates into a lateral displacement $\Delta r$. For small incidence angles $\gamma$, we find $\Delta r \approx f \gamma / n_{lens}$, i.e., the displacement $\Delta r$ increases in proportion to the focal distance $f$, which in turn is proportional to the collection radius $r_C$. The maximum tolerable lateral displacement $\Delta r_{max}$ is dictated by the finite detector size quantified by $r_D$ and by the focal spot size, e.g., quantified as second-moment-radius $w_0 = 2\sigma$ of a spot with an approximately Gaussian intensity distribution. In summary, we find

$$\gamma_{\max} \approx n_{\text{lens}} \frac{\Delta r_{\max}}{f} \propto \frac{\Delta r_{\max}}{a} \propto \frac{\Delta r_{\max}}{\sqrt{A_C}}, \qquad \Delta r_{\max} \approx r_D - w_0. \tag{9}$$

Thus, increasing the effective collection radius $r_C$ by scaling the lens size comes at the price of lowering the maximum tolerable angular misalignment $\gamma_{\max}$. In practice, lenses should hence be designed as big as the required angular alignment tolerance permits. In cryogenic systems, the achievable angular alignment tolerances might typically range from 0.2° to 2°, depending on the exact optomechanical implementation.

With these considerations, we can now outline a design procedure, which takes all limitations and trade-offs into account, see Fig. 2(b-e) and Appendix A for details. For a given refractive index $n$, we first consider the achievable spot size $w_0$, see Fig. 2b, where the vertical dashed line indicates the typical available refractive index of $n_{\text{lens}} \approx 1.5$ [22,23], which is a typical number for polymer photoresists at wavelengths in the visible and near-infrared spectral range. This spot size is dictated by the maximum ray angle of $\theta_{\max} \approx 48°$ according to Eq. (4) and by the fact that the Fresnel reflection at the lens surface increases with increasing distance from the optical axis. Figure 2b is based on Eq. (28) of Appendix A.2, which gives a more detailed description on how the spot size is extracted from the vectorial point-spread function and the associated distribution of the Poynting vector in the focal plane of the lens. Note that the consideration in Fig. 2b is independent of the absolute size of the lens. In a next step, we choose the minor half-axis $a$ of the lens as large as possible, given the limited angular alignment tolerance. To this end, we consider the ratio of the focal distance $f$ and the minor half-axis $a$, which is solely dictated by the refractive index $n$, see Fig. 2c. The upper limit for the focal distance can be estimated through Eq. (8) based on the known spot size $w_0$, see Fig. 2b, the known detector size $r_D$, and the required angular alignment tolerance as dictated by the application of the lens-equipped SNSPD. The lens gain $G_D$ is finally quantified by the effective collection radius $r_C$, see Eq. (2), which increases in proportion to the minor half axis $a$ of the lens and which, in addition, depends on the Fresnel reflection at the lens surface as dictated by refractive index $n_{\text{lens}}$, see Fig. 2d. Figure 2d is based on Eqs. (15) and (16) in Appendix A.1, which account for the position-dependent Fresnel reflection at the lens surface to calculate the effective collection area $A_C$ and the associated radius $r_C$.

Finally, we consider the case of an array of lensed detectors. For simplicity, we assume that the pitch of the detectors can essentially be chosen freely, thereby clipping the spheroidal lens surfaces along the perpendicular bisectors of the lines connecting the center points of neighboring detectors. This clipping removes the outer strongly inclined regions of the lens surfaces, which are subject to higher Fresnel reflections, such that strong clipping leads to an increased effective collection efficiency of the overall lens array. For a simplified calculation, we consider the case where the lens is clipped along a circular contour of radius $r_A$, centered about the optical axis, and we calculate the squared ratio of the effective collection radius $r_C$ and the geometrical contour radius $r_A$ for different refractive indices, see Fig. 2e. For $r_A \to a$, the value of $(r_C/r_A)^2$ corresponds to the one calculated in Fig. 2d, whereas for $r_A \to 0$, it approaches the Fresnel-limited power transmission in the case of normal incidence on a plane surface. We find that the value of $(r_C/r_A)^2$ does not increase significantly with decreasing $r_A$ as soon as $r_A < 0.7 \times a$, i.e., clipping the lens surfaces to even smaller contour radii $r_A$ does not lead to significant additional gain in the collection efficiency of the overall array. This estimate helps to determine the number of detectors and lenses that are needed or realize a high-fill-factor array with power transmission close to the Fresnel-limited value for the case of normal incidence on a plane surface. Note that typical photoresists used for multi-photon polymerization exhibit absorption of the order 1 dB/cm. For typical lens heights $h_0$ of less than 100 µm, this leads to absorption losses of less than 0.2 %, which can be neglected for most cases of practical interest. Note also that extended arrays of clipped lenses might also be efficiently produced by high-volume replication techniques such as nano-imprinting or hot embossing.

*3.2 Numerical simulations*

The spheroidal lens shape has been obtained in Section 3.1 based on simplified ray-optics considerations. For the clipped lens, however, additional side lobes of the point-spread function might occur, such that a wave-optical optimization could result in a slightly different optimum lens shape. In addition, the case of a slightly tilted illumination can only be analyzed in full using a wave-optical simulation. We therefore complement our design considerations by a numerical simulation of a specific lens design for a lensed SNSPD array. In the following, we consider a vacuum wavelength of $\lambda = 1550 \, \text{nm}$, a refractive index of $n_{\text{lens}} = 1.53$ and an SNSPD size of $A_D = 4.5 \times 4.5 \, \mu\text{m}^2$, which is consistent with the devices used for the experimental demonstration described in Section 4. We choose lenses with an apex height $h_0 = f = 70 \, \mu\text{m}$. For a spheroidal lens surface, Fig. 2a, the minor half axis of the spheroid is $a = 32.0 \, \mu\text{m}$, the effective collection radius amounts to $r_C = 30.0 \, \mu\text{m}$, and the effective lens gain is $G_D = 140$ for the unclipped lens, see Eqs. (2),(8) and Figs. 2c,d. We further estimate an achievable spot size radius of $w_0 = 0.80 \, \mu\text{m}$, see Fig. 2(b). This leads to a maximum allowed lateral displacement of $\Delta r_{\max} = 1.45 \, \mu\text{m}$ for a detector size of radius $r_D = 2.25 \, \mu\text{m}$, corresponding to a maximum illumination tilt of $\gamma_{\max} = 1.8°$, see Eq. (9), which can be well achieved in a fixed cryogenic setup without any means for further adjustment during the experiment.

Based on this design we, then investigate the behavior of densely packed lenses as part of an array. Naturally, arranging lenses in a gapless array requires some kind of clipping, depending on the structure of the underlying lattice. For the clipped lens, additional side lobes of the point-spread function might occur, and the spheroidal refracting surface of the lenses might not any more lead to the maximum possible concentration of incident optical power into the active area of a commonly rectangular SNSPD. To investigate this effect, we numerically optimize the refracting surfaces of clipped lenses and compare the resulting shapes and collection efficiencies to the ones of spheroidal surfaces. For the numerical optimization, we use an in-house simulation software written in Python and running on a graphic processing unit (GPU). The software uses the scalar wide-angle unidirectional wave-propagation method for step-index structures proposed in [24]. This allows for fast and realistic wave-optical modeling of micro-optical components beyond the thin-element approximation. For this method, exceptional consistency with rigorous finite-difference time-domain (FDTD) solutions of Maxwell's equations has been shown in terms of focal intensity distributions [24], while the underlying calculations are considerably faster than those associated with various wide-angle beam-propagation-methods. Note that the field computation could be further accelerated by exploiting the rotational symmetry of the problem in case the incident plane wave propagates along the optical axis [25].

For a simple implementation of the design procedure, we assume a rotationally symmetric lens with a fixed apex height of $h_0 = 70 \, \mu\text{m}$, see Fig 1b, and parametrize the height of the refracting surface by a polynomial $h(r) = h_0 + c_2 r^2 + c_4 r^4 + ...$ as a function of the lens radius $r$. In a first step, we further assume clipping along a circular contour $r_A = 22.4 \, \mu\text{m}$, corresponding to a clipping ratio of $r_A / a = 70\%$ for the spheroidal lens shape. According to Fig. 2e, this choice of the clipping radius should allow for a lens array with an overall collection efficiency that is close to its theoretical optimum dictated by Fresnel losses at normal incidence. For the optimization, we consider a single free-standing lens, which is illuminated by a plane wave incident along the optical axis of the lens, and we optimize the polynomial coefficients $c_k$ for maximum power in the detector area $A_D = 4.5 \times 4.5 \, \mu\text{m}^2$. We find that, when using only two free coefficients $c_2$ and $c_4$, the numerical optimization of the clipped lens surface leads to a rather marginal improvement of 0.1 % with respect to the reference case of a clipped

spheroid – the numerical values of the coefficients are specified in the third row of Table 1 below (clipped polynomial, optimized as single lens with $r_A = 22.4\,\mu\text{m}$). For three free coefficients $c_2$, $c_4$, and $c_6$, this improvement increases to 0.3 %. We may hence conclude that lens arrays on top of SNSPD may indeed be designed by merging simple spheroids, without the need for further numerical optimization.

In a second step, we extend the simulation to an entire hexagonal array of lens-equipped SNSPD with spacing $2r_A$, see inset of Fig. 3a. For this array, we chose again $r_A/a = 70\,\%$. The white line in Fig. 3a indicates the shape of the underlying unclipped spheroid, and the green line refers to the contour of the previously numerically optimized lens surfaces with only two free coefficients $c_2$ and $c_4$. The two shapes hardly differ, which is consistent with the fact they result in essentially the same performance. The colors in Fig. 3a refer to the electric-field magnitude which is depicted in the (*x*, *z*)-plane. The blue dashed rectangle in the inset refers to the computational area, for which we use periodic boundary conditions to mimic an infinitely extended lens array. From the simulation, we also extract the intensity distribution along the *x*-axis in the focal plane, both for normal and for slightly angled incidence, see Fig. 3b. The solid black line refers to the simulated intensity profile obtained for the clipped lens array, and the blue gives the profile obtained for the full unclipped spheroid, both for normal incidence. As expected, the clipping leads to a broadening of the intensity profile – the second-moment-radius of the intensity distribution of the clipped lens amounts to $w_{0,c} = 2\sigma_c = 1.1\,\mu\text{m}$, whereas a second-moment-radius of $w_{0,nc} = 2\sigma_{nc} = 0.7\,\mu\text{m}$ is found for the non-clipped spheroid lens. The simulated value for the non-clipped lens is in reasonable agreement with the value $w_0 = 0.8\,\mu\text{m}$ estimated based on Fig. 2b and Appendix A.2. We attribute the slight differences mainly to the approximations related to the position-dependent Fresnel losses in both techniques. Note that even for the clipped lens, side lobes of the intensity distribution do not play a significant role. The area shaded in blue in Fig. 3(b) indicates the actual width of the SNSPD. We further simulate the intensity distributions for a direction of incidence that deviates from the optical axis by tilt angles $\gamma$ of 3°, 6°, and 9°, see dotted curves in Fig. 3b. The dashed vertical lines correspond to the associated lateral offsets $\Delta r$ of these intensity distributions, obtained by assuming a focal length of $f = h_0 = 70\,\mu\text{m}$ and a linear relationship $\Delta r \approx f\gamma/n_{\text{lens}}$. Note that this linear relationship is only valid in the limit of small angular deviations $\gamma < 5°$ and that the maximum of the intensity distribution clearly deviates from the respective dashed line for tilt angles $\gamma$ of 6°, and 9°.

We further numerically calculate the effective gain $G_D$ of an individual clipped lens as a function of the tilt angle, see Fig. 3c. To this end, we integrate the intensity in the focal plane of the lens over the active area $A_D = 4.5 \times 4.5\,\mu\text{m}^2$ of the SNSPD, which leads to an almost constant lens gain $G_D \approx 77$ for tilt angles $\gamma \leq 1.5°$ with a 1 dB decay at $\gamma_{\text{1dB}} = 2.5°$. This result is in reasonable agreement with maximum illumination tilt of $\gamma_{\max} = 1.8°$ estimated for the corresponding ideal un-clipped lens. Note that the lateral extension of the dashed intensity distributions in Fig. 3c do not change strongly with tilt angle $\gamma$. The lens array can hence be deliberately designed for reception of light from directions that slightly deviate from normal incidence by simply introducing a lateral offset between the SNSPD and the optical axis of the corresponding lens. Similarly, a single lens may be combined with multiple SNSPD that are directly adjacent to each other to increase the maximum tolerable angular misalignment or to enable angle-resolved reception of incoming signals. We finally calculate the effective fill factor $\eta$ of the hexagonally arranged lensed detectors. Each lens covers a hexagonal cross-section area of $A_L = 2\sqrt{3} \times r_A^2$. Thus we find

$$\eta = \frac{A_C}{A_L} = \frac{A_C}{A_D}\frac{A_D}{A_L} = G_D \frac{A_D}{A_L} \approx 77 \times \frac{A_D}{2\sqrt{3} \times r_A^2} \approx 89\,\% \ . \tag{10}$$

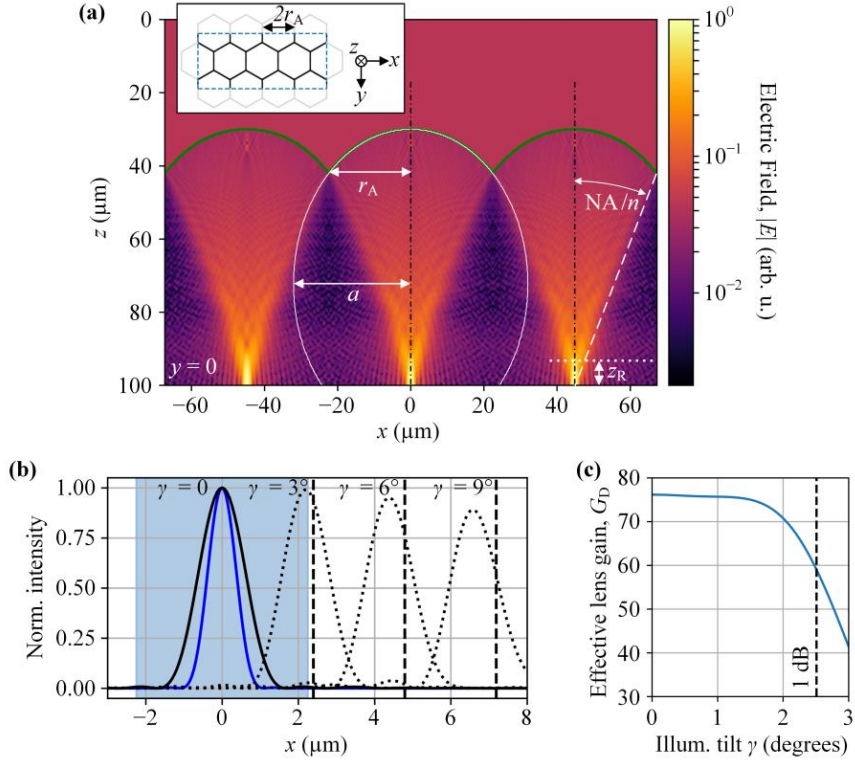

**Fig. 3.** Simulation of an arrangement of densely packed pillar-shaped microlenses with hexagonal cross-section using the wave-propagation-method [24]. Each SNSPD has a detector area of $A_D = 4.5 \times 4.5\,\mu m^2$, and the apex height of the lenses is fixed to $h_0 = 70\,\mu m$. The arrangement is illuminated by a plane wave from the top. For designing the surface shape, we first consider individual free-standing lenses with circular cross-section (clipping radius $r_A = 22.4\,\mu m$), which could be inscribed into the hexagonal pillars, see Inset Fig. 3a. We numerically optimize the lens shape such that maximum power is collected by the detector. (a) Cross section through the hexagonal microlens array with spacing $2r_A$. The green contour lines show an optimized polynomial lens surface with two free coefficients $c_2$ and $c_4$. For comparison, the white line shows the spheroidal surface contour with the same height (minor half-axis of ellipsoid $a$, $r_A/a = 70\%$ Fig. 2e). The colors refer to the electric-field magnitude. The asymptotic divergence angle $\beta = 0.37$ (corresponding to $21°$) inside the lens is given by $n \sin \beta = NA$, leading to a (one-sided) depth of field of $z_{DOF} = (1.77/2)(\lambda/n)/(NA/n)^2 = 7\,\mu m$ according to [31,32]. The blue dashed rectangle in the inset designates the computational region. We use periodic boundary conditions to mimic an infinitely extended lens array. (b) Intensity distribution along the $x$-axis on the chip surface, both for normal (solid lines) and for slightly angled incidence (dotted lines). The solid black line refers to the normalized intensity profile obtained for the hexagonal lens array with optimized polynomial surface, and the solid blue line gives the intensity profile for a free-standing unclipped spheroid, both for normal incidence of a plane wave. The blue shaded area indicates the width of the SNSPD. The dotted curves are the normalized intensity distributions for a plane wave incident with tilt angles $\gamma$ of $3°$, $6°$, and $9°$ measured towards the surface normal of the substrate. The dashed vertical lines correspond to the associated lateral offsets $\Delta r \approx f \gamma / n_{lens}$ of the focal spot as estimated by geometrical optics using a focal length of $f = h_0 = 70\,\mu m$. (c) Expected effective lens gain $G_D$ as a function of the tilt angle. The dashed line indicates the 1 dB decay, which occurs at an illumination tilt angle $\gamma$ of approximately $2.5°$.

This is more than two times higher than the best effective fill factor of $\eta = 36\%$ that was previously reported for an SNSPD array consisting of 1024 individual detectors that cover on an area of 1.6 mm × 1.6 mm [26]. Note that a single detector of this array has a size of 30 µm × 30 µm and a nanowire length larger than 3 mm – much larger than the 100 µm used for the SNSPD in our experiment.

To provide an overview and a comparison of the different aspherical lens surfaces considered in this section, we summarize them in Table 1 together with the respective lens gain $G_D$ and spot-size radius $w_0$, that can be expected from a lens arranged into a hexagonal array with spacing $2r_A = 2 \times 22.4\,\mu m$, see inset of Fig. 3a. As a reference, we consider a spheroid surface that is clipped to the hexagonal contour dictated by the array, second row in Table 1 ("Clipped spheroid"). As a very simple alternative, we consider a spherical surface that is clipped to the same contour, third row ("Clipped sphere"). The radius of curvature of this surface is chosen to provide maximum lens gain for a clipping along a circular contour with radius $r_A = 22.4\mu m$. These shapes are then benchmarked against a clipped polynomial surface with two free coefficients $c_2$ and $c_4$, again optimized for best coupling under circular clipping with radius $r_A = 22.4\mu m$, fourth row. For comparison, we also specify the coefficients $c_2$ and $c_4$ of a polynomial surface that leads to the smallest sum of squared deviations from the spheroidal and the spherical surface within a circular aperture of radius $r_A = 22.4\mu m$, indicated in parentheses in the third and the fourth columns of Table 1. We further validate the designs by comparison to a clipped polynomial surface, again with two free coefficients $c_2$ and $c_4$, which has been optimized for highest lens gain within the entire hexagonal array, fifth row. Interestingly, the resulting lens gain of $G_D = 76.3$ is even slightly worse than the gain $G_D = 76.5$ obtained for the simple spheroidal shape. We attribute this to the limitations of two-coefficient polynomial in representing the ideal surface, which becomes more apparent for larger apertures. For practically relevant use cases, however, these deviations are insignificant. In the fifth column, we specify the maximum deviation of the respective surface to the ideal spheroid shape, measured parallel to the optical axis. We find that the spherical shape shows the largest deviation to the ideal spheroid shape, which ranges up to 0.45 µm. For the other considered lens shapes, the deviations are smaller, and the values for the lens gain $G_D$ as well as the achievable spot sizes $w_0$ differ from the optimum spheroid only insignificantly.

**Table 1.** Comparison of lenses with various shapes of the refracting surface. All lenses have an apex height of $h_0 = 70\mu m$ above the substrate. The coefficients $c_2$ and $c_4$ in the third and the fourth columns refer to a parametrization of the radius-dependent height above the substrate of the form $h(r) = h_0 + c_2 r^2 + c_4 r^4 + ...$. For the spheroidal and the spherical shape, $c_2$ and $c_4$ refer to a polynomial surface that leads to the smallest sum of squared deviations from the respective surface within a circular aperture of radius $r_A = 22.4\mu m$. The last two columns refer to the spot radius $w_0$ and the lens gain obtained in a hexagonal a hexagonal array with spacing $2r_A = 2 \times 22.4\mu m$, see inset of Fig. 3a.

| Lens shape | Optimized as | Clipping radius $r_A$ = 22.4 µm | | | Hexagonal array | |
|---|---|---|---|---|---|---|
| | | $c_2$ (µm$^{-2}$) | $c_4$ (µm$^{-4}$) | Max. deviation to spheroid (µm) | $w_0$ (µm) | $G_D$ |
| Clipped spheroid | — | (-0.0202) | ($-7.36 \times 10^{-6}$) | — | 1.118 | 76.5 |
| Clipped sphere | Single lens ($r_A$ = 22.4 µm) | (-0.0176) | ($1.16 \times 10^{-5}$) | 0.45 | 1.182 | 72.1 |
| Clipped polynomial | Single lens ($r_A$ = 22.4 µm) | -0.0193 | $-9.34 \times 10^{-6}$ | 0.12 | 1.120 | 76.1 |
| Clipped polynomial | Hexagonal array | -0.0190 | $-9.10 \times 10^{-6}$ | 0.23 | 1.143 | 76.3 |

## 4. Experimental demonstration

### 4.1 Device fabrication

To prove the practical viability of our approach, we fabricated a pair of SNSPD from a magnetron-sputtered, 5 nm-thick niobium nitride (NbN) film on a sapphire substrate, see [19,27] for details of the fabrication process. For the experiment, two identical 4.5 µm × 4.5 µm detectors, designed for DC operation, were structured 150 µm apart from each other in the center of a 3 mm × 3 mm chip, see Fig. 4a. One detector of the pair is used with a 3D-printed microlens ("lensed detector"), while the other is left blank ("reference detector"). The meandered nanowires in both detectors are 110 nm wide, 105 µm long, and cover the detector area with a fill factor of about 50%. Insets 1 and 2 of Fig. 4a show scanning-electron-microscope (SEM) images of the lens and the meandered SNSPD, respectively. For passivation, the nanowires are covered with a 20 nm-thick aluminum-nitride (AlN) layer to prevent oxidation, see Inset 3 of Fig. 4a. The critical temperature of the samples is $T_c = 12$ K, achieved by tuning the stoichiometry of the NbN film. Together with the patterning technique used [19], this leads to comparatively high values of the so-called switching currents $I_{sw}$, at which the devices switch from superconducting to normal state. For the two SNSPD used in our experiment, we find a switching current $I_{sw}$ of 51 µA for the lensed device and of 55 µA for the device without lens. The 3D-printed microlenses were fabricated from negative-tone photoresist (VanCoreA, Vanguard Automation GmbH, $n = 1.53$ at 1550 nm) by multi-photon lithography [20,28,29] using an in-house-built lithography system with a 63x/1.4 objective (Zeiss Plan-Apochromat 63x/1.4 Oil DIC M27), galvanometer-actuated mirrors and a 780 nm femtosecond laser (Menlo C-Fiber 780 HP, pulse width 58 fs). We use a numerically optimized lens design based on a fourth-order polynomial $h(r) = h_0 + c_2 r^2 + c_4 r^4$ which very well approximates a theoretically optimum spheroidal shape with a minor half axis of $a = 32.0\,\mu m$ and a focal length of $f = 70\,\mu m$, see Section 3.2 for details of the design. Since our experimental validation is limited to a free-standing lens, we chose a slightly larger clipping radius of $r_A = 30\,\mu m$, for which we expect a lens gain $G_D = 117$ along with a 1 dB decay at a tilt angle of $\gamma_{1dB} = 2.9°$.

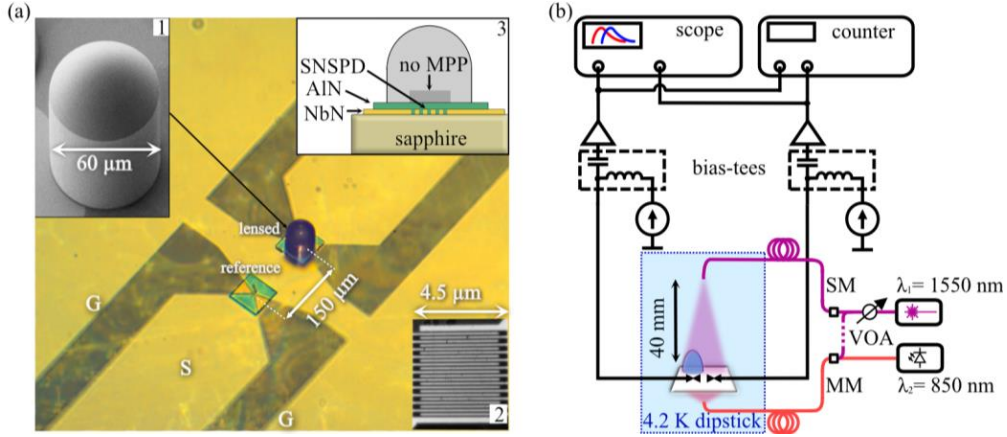

**Fig. 4.** Experimental demonstration using a pair of SNSPD on the same chip. One detector is equipped with a 3D-printed lens ("lensed detector") while the other is left blank ("reference detector"). **(a)** Optical microscope image of the dual-detector SNSPD chip. The nanowires are patterned into a 5 nm-thick NbN layer on a sapphire substrate and passivated with a 20 nm-thick layer of AlN. Detectors are biased and read out using coplanar waveguide transmission line with ground (G) and signal (S) electrodes patterned in the same NbN layer. The scanning electron microscope (SEM) images in Insets 1 and 2 show the lens and the meandered SNSPD (4.5 µm × 4.5 µm), respectively. Inset 3 shows a schematic cross section of a lensed detector. To avoid degradation of the nanowire during lens fabrication by multi-photon polymerization (MPP), a small cylindrical volume ("no MPP", dark grey) is left unexposed during the lithography. The material in this region is then solidified by UV flood exposure after development of the structure. **(b)** Experimental setup for characterizing the dual-detector SNSPD chip. The chip is mounted in a dipstick tube, and both detectors are connected to individual coaxial cables, which are used for biasing through a pair of bias tees and for reading out the electrical pulses from the SNSPD. The pulses are amplified and fed to a counter and a real-time oscilloscope. The optical test signal is derived from a fs-laser with an emission wavelength of 1550 nm, which is coupled to a subsequent variable optical attenuator (VOA). The SNSPD are front-side illuminated by the open end of a single-mode (SM) fiber that is approximately 40 mm away from the chip surface. This leads to an approximately equal plane-wave-like illumination of both devices. The backside can be illuminated via a multimode (MM) fiber with a continuous-wave laser source having a wavelength of 850 nm or with the 1550 nm femtosecond laser, see dotted line. The facet of the MM fiber is fixed 3 mm from the chip's backside. The dipstick is gradually cooled down to 4.2 K in liquid helium.

For fully automated 3D lithography, we use markers in the direct vicinity of each detector along with techniques for detection of the chip height and tilt. This leads to a lateral and vertical alignment precision of the order of 100 nm [30]. Our lithography process produces approximately spheroidal voxels with axes of about 0.5 µm × 0.5 µm × 1.6 µm, where the longest dimension is oriented along the illuminating beam axis of the lithography system. In addition, the photoresist features an isotropic linear shrinkage of less than 1 %. These effects are known and can be compensated in the design, leading to an overall precision of the total structure height and consequently of the axial position of the refracting surface significantly better than 1 µm. Besides that, the lens surface might be subject to shape deviations that we measure by white-light interferometry. We find that the typical deviation of actually printed surfaces from their respective theoretical shape is below 100 nm over the entire aperture and that the typical root-mean-square (RMS) surface roughness is of the order of 40 nm [28]. For estimating the impact of these inaccuracies on the lens performance, we first separate the influence of the shape of the refracting surface from the influence of its axial position, directly linked to the lens height. The one-sided depth of field $z_{\mathrm{DOF}} = (1.77/2)(\lambda/n)/(\mathrm{NA}/n)^2$ [31,32] of the focused beam inside the polymer lens is $z_{\mathrm{DOF}} = 7\,\mu\mathrm{m}$ for $\lambda = 1.55\,\mu\mathrm{m}$, $n = 1.53$, and $\mathrm{NA}/n = 0.37$, see horizontal dotted line in Fig. 3(a). Assuming a maximum height deviation of 1 µm, we estimate a deterioration the coupling efficiency to the SNSPD by approximately 0.2 %, which is of no practical relevance. The tolerable shape deviation from the optimum spheroidal lens surface cannot be estimated as easily, because the impact on the coupling efficiency depends on the exact type of the associated aberration. For an order-of-magnitude estimate, we compare the expected shape deviations to those that occur between the ideal spheroidal surface and its spherical approximation as specified in the second and the third row of Table 1. For the circular aperture radius of $r_{\mathrm{A}} = 22.4\,\mu\mathrm{m}$ considered in Table 1, the maximum deviation between spheroidal surface and its spherical approximation amounts to 450 nm, while the lens gain deteriorates only slightly – from and initial value of $G_{\mathrm{D,spheroid}} = 76.5$ of the ideal spheroid to $G_{\mathrm{D,sphere}} = 72.1$ for the spherical approximation, corresponding to a reduction of approximately 6 %. Since the systematic shape deviations due to fabrication tolerances amount to only 100 nm, the impact on the lens gain should be much smaller. Similarly, surface roughness with a root-mean-square deviation of 40 nm can be expected to have no significant influence on the overall detector performance. The lenses were written conservative writing parameters, without any special acceleration techniques, leading to rather high printing times of approximately 20 min per lens. We expect that this time can be greatly reduced by optimized writing techniques. We found the printing processes to be very reliable, once the correct printing parameters have been found. In the course of our experiments, we printed multiple chips with the same set of parameters, comprising more than 30 lenses overall, which were all fully functional.

The microlenses can be operated over a broad wavelength range. For the currently used resist materials, absorption is typically negligible down to wavelengths of approximately 500 nm [23]. The transparency range can be extended further down to 300 nm by using suitable photo initiators [33]. The slightly higher refractive index $n_{\mathrm{lens}} = 1.58$ at 300 nm does not have any significant effect on the effective lens gain according to our simulations. Note that the SNSPD might experience significant degradation when directly exposed the focused laser light of the lithography system. In such cases, we observed that the room-temperature resistance of the device increases by factor of more than three. At the same time, the critical temperature of the superconducting transition in the nanowire is found to be reduced to (6…7) K, and the critical currents fall below 10 µA. To avoid this degradation we use a technique similar to the on reported in [20], leaving a small cylindrical volume with diameter of 12 µm and height of 3 µm unexposed during lithography, see Inset 1 of Fig. 4a. The fabricated structures are developed in propylene-glycol-methyl-ether-acetate (PGMEA), flushed with isopropanol, and subsequently blow-dried. A post-exposure with UV light (EFOS Novacure N2000, 500 mW/cm² for 40 s) [20] solidifies the liquid resist, which is encapsulated in the vicinity of the meandered nanowire. Note that this UV dose is rather low in comparison to those reported in [23]. This might be attributed to the rather small volumes of the 3D-printed microlenses (maximum apex height of $h_0 < 100\,\mu\text{m}$) in comparison to the 2 mm-thick layers investigated in [23] and to the fact that small curing-induced changes of the refractive index as observed in [23] are not crucial for the functionality of our structures.

*4.2 Experimental setup*

The experimental setup for characterizing the fabricated pair of SNSPD is shown in Fig. 4b. The detectors are directly connected to individual 50 Ω coplanar on-chip transmission lines for readout and biasing, see Fig. 4a. For introducing the sample into the cryostat, the chip is attached to a submount comprising an adapter plate for two coaxial cables. The assembly is inserted into a vacuum-tight dipstick tube with helium (He) contact gas at a pressure of 10 mbar at 300 K. The dipstick is gradually immersed in a liquid-$^4$He transport dewar and reaches 4.2 K within approximately 30 min. Similar to [29,34,35], the printed lenses proved to be stable during repeated cooldown / warm-up cycles in a temperature range from 300 K down to 4.2 K – we performed around 10 cycles without observing any lens detachment or peeling-off. For testing, light is supplied to the device by a pair of fibers inside the dipstick tube and emitted towards to the front and the back surface of the chip, see Fig. 4b. For front-side illumination, we use a standard single-mode (SM) fiber (Thorlabs SMF-28-J9, Hytrel jacket with 900 µm diameter), which is fed by a pulsed femtosecond laser emitting at a wavelength of 1550 nm with a repetition rate $f_{\text{rep}} = 100\,\text{MHz}$ and a pulse duration of approximately 150 fs. The fiber ends about 40 mm above the center of the chip, which leads to an approximately equal plane-wave-like illumination of both SNSPD that are spaced by only 150 µm with a lens of 60 µm diameter on top of one of the devices. The backside of the chip can be illuminated via a multimode (MM) fiber using either the 1550 nm femtosecond laser, see dotted purple line in Fig. 4b, or a continuous-wave laser with an emission wavelength of 850 nm, solid red line. The facet of the MM fiber is fixed 3 mm from the chip's backside. A DC bias is applied to the SNSPD through coaxial cables using a pair of bias-tees. Voltage pulses from the detectors are transmitted through the RF branches of the bias-tees, amplified by room-temperature amplifiers (MITEQ AFS4), and finally detected by a real-time oscilloscope (Keysight Infiniium, 33 GHz acquisition bandwidth) and a pulse counter (SRS SR620). Using the real-time oscilloscope, a reset time of both detectors of $t_{\text{res}} \approx 2\,\text{ns}$ is measured from the 90 % to 10 % fall times. In this measurement, we evaluated 10 000 pulses and averaged the individual fall times. The reset time is smaller than the repetition period of the fs-laser, such that no detrimental impact on the pulse counting rate is to be expected.

*4.3 Count-rate measurements*

We measure the photon count rates PCR$_{\text{lens}}$ and PCR$_{\text{ref}}$ of the lensed and the reference detector at several average incident optical powers of the femtosecond laser. To extract the lens gain $G_D$, we need to assume equal absorption efficiencies ABS and intrinsic detection efficiencies IDE for both detectors, see prerequisites of Eq. (2). Because the detectors are made from the same film, have the same geometry and orientation of the nanowires, and are placed close to each other, it is a safe to assume equal ABS. Regarding intrinsic detection efficiency IDE, we have to account for its dependence $\text{IDE}(\lambda, I_b/I_{\text{sw}})$ on both the wavelength $\lambda$ and the relative bias current $I_b/I_{\text{sw}}$, see Appendix B for an exemplary behavior of similar detectors. To experimentally adjust for similar intrinsic detection efficiencies IDE, we first measure the dependencies of the pulse count rate $\text{CR}(I_b/I_{\text{sw}})$ on the relative bias current for both detectors under pulsed back-side illumination with a wavelength of 1550 nm, see Fig. 5a. We find that both devices exhibit nearly the same behavior such that operating them at the same relative bias current $I_b/I_{\text{sw}}$ leads to the same IDE. In our measurement, we found similar absolute values of the count rates CR, which means that both devices are subject to the same flux of incoming photons and should hence feature the same rate $R_a$ of absorbed photons. We collectively fit the measured data points of both detectors with a theoretical model function that is adapted from Eq. (8) and (9) in [36],

$$\mathrm{PCR}(I_b/I_{sw}) = R_a \times \mathrm{IDE}(I_b/I_{sw}) = \frac{R_a}{2} \mathrm{erfc}\left[ q\left(1 - \frac{I_b/I_{sw}}{I_{0.5}/I_{sw}}\right) \right]. \qquad (11)$$

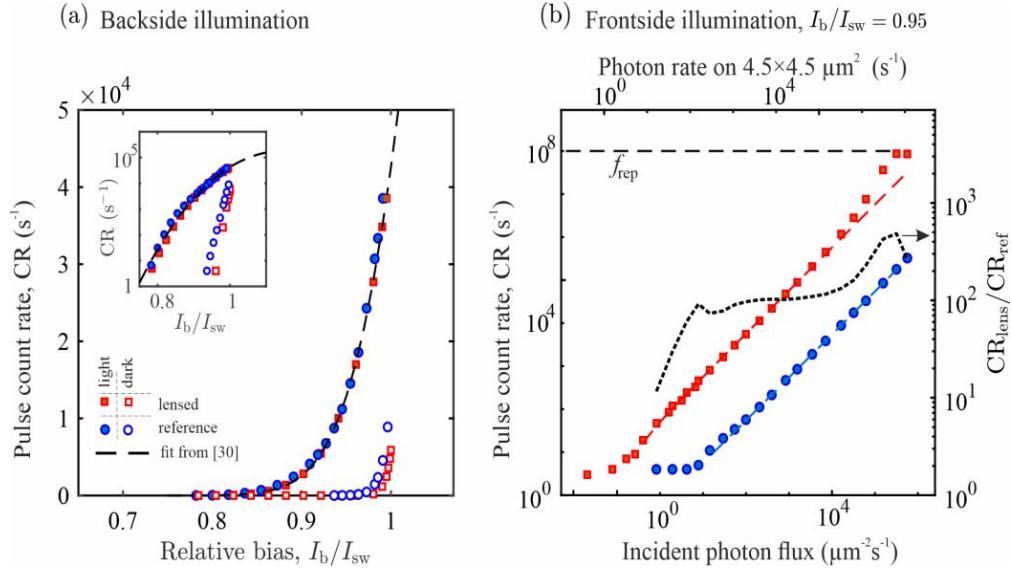

**Fig. 5.** Experimental results for the dual SNSPD chip with lensed detector (■, □) and with reference detector (●, ○), illuminated with light at a wavelength of 1550 nm. **(a)** Dependence of pulse count rates (CR) on the relative bias current $I_b/I_{sw}$. The closed markers (■, ●) show the data together with a fit (– – –) according to Eq. (11) [36], when illuminating the detectors from the back side. The measurements are taken at an estimated rate of approximately $4.8 \times 10^5$ photons per second that are incident on the 4.5 µm × 4.5 µm area covered by the meandered nanowire of the SNSPD. The open markers (□, ○) show the dark count rates (DCR), measured with a metal cap on the FC/APC connector of the vacuum feedthrough to block any stray optical photons. The DCR are negligible for bias currents $I_b < 0.98\, I_{sw}$. The CR are plotted on a logarithmic scale in the inset. **(b)** Double-logarithmic plot of the pulse count rate CR for frontside illumination. The lower horizontal axis indicates the incident photon flux, which may be translated into a rate of photons incident on the reference detector with an active area of 4.5 µm × 4.5 µm, see upper horizontal axis. For the lensed detector, the rate of captured photons is increased according to the lens gain. The dashed lines are fits of the form $\mathrm{PCR} = c_1 P_{opt}$ with parameter $c_1$ resulting from a fit to the central portion of the curves. At high photon rates, the CR dependence of the lensed SNSPD is super-linear, which might be attributed to the multi-photon bolometric regime (MBR) [41]. The horizontal dashed line shows the maximum CR dictated by the repetition rate $f_{rep}$ of the femtosecond laser. Right-hand axis: Ratio $\mathrm{CR}_{lens}/\mathrm{CR}_{ref}$ of measured pulse count rate CR. At the plateau in the single-photon regime, we find an effective lens gain $G_D \approx 100$, which is in reasonable agreement with a simulated value of $G_D = 117$.

In this relation, erfc denotes the complementary error function, and $R_a$ is the rate of absorbed photons. $I_{0.5}$ refers to the so-called inflection current at which the IDE amounts to 50 %, and $q$ is a dimensionless parameter. Assuming identical rates $R_a$ of absorbed photons and identical parameters $I_{0.5}$, $I_{sw}$ and $q$ for both devices, the best fit of the measured PCR is obtained for $q = 10.5$, $I_{0.5}/I_{sw} = 1.063$ and $R_a = 2.16 \times 10^5$ s$^{-1}$. From the plot in Fig. 5a, we find that the PCR does not reach a plateau within the range of usable bias currents $I_b < I_{sw}$. This indicates that the device is operated in its non-saturated regime [37] with $\mathrm{IDE} \ll 1$, as expected for NbN SNSPD at 4 K [38]. To further support the notion that identical relative bias currents $I_b/I_{sw}$ lead to similar IDE for both detectors, we illuminate the devices from the backside with alternating wavelengths $\lambda_1 = 850$ nm and $\lambda_2 = 1550$ nm and compare the associated PCR, see Appendix B for details.

For the lens gain measurement with frontside illumination, Fig 5b, we operate the detectors at a fixed relative bias $I_b/I_{sw} \approx 0.95$ and sweep the incident optical power. Sweeping the optical power rather than the bias current allows to isolate the effect of the lens from potential distortions of the measurement results due to bias-dependent non-uniformities of the detection efficiency along the tightly bent meandered nanowires, which may be subject to current-crowding effects at low bias currents [39]. The relative biases $I_b/I_{sw} \approx 0.95$ were chosen to ensure stable operation without excessive impairments by dark counts.

To determine the incoming photon flux in the frontside illumination experiment, we first measure the optical power in the SMF at the input to the dipstick. Accounting for the optical losses subsequent fiber assembly, we can then estimate the optical power radiated towards the detector chip and the optical intensity on the chip surface. To this end, we assume a diverging Gaussian beam having its waist at the output facet of the illuminating SMF, 40 mm away from the chip surface. This intensity is then translated into the incident photon flux, see horizontal axis at the bottom of Fig. 5b. We also calculate the rate of photons that are incident on the 4.5 µm × 4.5 µm area of the reference detector, see upper horizontal axis of Fig. 5b. Note that we did not directly measure the photon flux associated with the backside illumination experiment in Fig. 5a, since the losses of the underlying multi-mode fiber (MMF) assembly were not exactly known. We may, however, estimate the flux associated with Fig. 5a from the data shown in Fig. 5b. Specifically, we find a measured pulse count rate CR of $1.3 \times 10^4$ s$^{-1}$ at a bias of $I_b = 0.95\,I_{sw}$ in the back-side illumination experiment, Fig. 5a. According to Fig. 5b, this pulse count rate can be associated with a photon rate of $4.8 \times 10^5$ photons per second, incident on the 4.5 µm × 4.5 µm area of the reference detector.

To evaluate the lens gain, we need to check whether the detectors operate in the single-photon regime. At the input aperture of the detection system, the laser pulses have an average power $P_{opt}$, corresponding to an average number of $m = P_{opt}/(f_{rep}\hbar\omega)$ photons per impulse, where $\hbar$ is the reduced Planck's constant and $\omega$ is the angular frequency of the light. For each laser pulse, the probability to observe a voltage pulse originating from $n$ absorbed photons can be estimated from Mandel's formula [40] for light with Poisson statistics,

$$p_m(n) = \frac{(\text{SDE} \times m)^n}{n!} e^{-\text{SDE} \times m}. \qquad (12)$$

The total photon detection probability $p_{m,\text{tot}}$, i.e, the probability to detect at least one photon per impulse, can then be expressed and approximated for the case of very high and very low average photon numbers $m$,

$$p_{m,\text{tot}} = \sum_{n=1}^{\infty} p_m(n) \approx \begin{cases} \text{SDE} \times m \approx p_m(1) & \text{for SDE} \times m \ll 1 \\ 1 & \text{for SDE} \times m \gg 1 \end{cases}. \qquad (13)$$

For small average detected photon numbers $\text{SDE} \times m \ll 1$, the detection probability $p_{m,\text{tot}}$ is approximately equal to the probability $p_m(1)$ to detect exactly one photon and approximately equal to the detected average number of photons per impulse. In this case, the photon count rate (PCR) is approximately equal to the observed count rate (CR) of voltage pulses. For large detected average photon numbers, the detection probability approaches unity, and the CR of the voltage pulses approaches the repetition rate $f_{rep}$ of the laser. The measured average count rate of voltage pulses depends on the detection probability $p_{m,\text{tot}}$ and on the repetition frequency of the laser pulses,

$$\text{CR} = p_{m,\text{tot}}\, f_{rep}. \qquad (14)$$

Figure 5b shows the measured average CR as a function of the incident photon rate for both the lensed (■) and for the reference detector (●). Obviously, the average count rate CR of voltage pulses cannot exceed the laser repetition rate $f_{\text{rep}}$, indicated by a horizontal dashed line. Except for the case of the lensed detector and the highest optical powers, even $\text{CR} \ll f_{\text{rep}}$ holds, and thus the case $\text{SDE} \times m \ll 1$ in Eq. (13) applies. In this case, the linear relation between $p_{m,\text{tot}}$ and $m$ from Eq. (13) is indeed seen in Fig. 5b as a linear dependence of CR on the incident photon rate, indicated by a line with unity slope in the double-logarithmic plot over most of the measurement range. For the chosen relative biases of $I_b/I_{\text{sw}} \approx 0.95$ we find a dark count rate for both detectors of $\text{DCR} \approx 5\,\text{s}^{-1}$. This leads to the deviation from the linear dependence at low incident photon rates. For the lensed detector, a super-linear dependence is observed at high incident photon rates, before the pulse count rate CR finally saturates approaches the repetition rate $f_{\text{rep}}$ of the laser. We attribute this behavior to the multi-photon bolometric regime (MBR) [41], which is caused by simultaneous absorption of multiple photons within a region comparable to the mean hot-spot size of the nanowire. These multi-photon-generated hot-spots have a much higher probability of switching the SNSPD from the superconducting to the normal state than their single-photon-generated counterparts, thereby leading to a higher intrinsic detection efficiency IDE and thus a higher SDE in Eq. (12). As expected, multi-photon bolometric events are more likely to happen for the lensed device because of the increased optical intensity, whereas they are not observed for the reference detector. In addition, we display the ratio $\text{CR}_{\text{lens}}/\text{CR}_{\text{ref}}$ in Fig. 5b, see axis on the right-hand side. In the single-photon regime, i.e., for medium optical input powers, where neither the DCR nor the MBR plays a role, this ratio exhibits a plateau, which corresponds to the effective lens gain $G_D$. The obtained value $G_D \approx 100$ is in reasonable agreement with the simulated value of $G_D = 117$.

## 5. Discussion

We have demonstrated that SNSPD with 3D-printed light-collection lenses can overcome the design conflict between large collection area and short nanowire length. This applies not only to illumination through free-space plane waves, but also to coupling of SNSPD to optical fibers. Specifically, 3D-printed lenses allow to reduce the spot size of the focused light to a second-moment spot radius of $w_0 \approx 0.5\lambda$. At a wavelength of $\lambda = 1.55\,\mu\text{m}$, it is hence possible to reduce the detector area to approximately $2\,\mu\text{m} \times 2\,\mu\text{m}$, when disregarding any angular alignment tolerance. This area could be covered with a fill factor of FF = 50 % by using a nanowire with a typical width of 0.1 µm and a length of only 20 µm. This is much shorter than the nanowire length of 1.7 mm used in a previous demonstration of a fiber-coupled SNSPD [42] in which the detector area was 15 µm × 15 µm, slightly bigger than the size of a SMF core with a typical diameter of 10 µm. Nanowires as short as 20 µm allow for even shorter reset times than experimentally demonstrated in this work, enabling maximum photon count rates in the GHz range [5]. Our design considerations show that, in case of illumination by free-space plane waves, numerically optimized surfaces do not offer a significant advantage over idealized spheroid surfaces, even for clipped lenses that arrange in densely packed hexagonal arrays. The shape of the refracting surfaces can be simply derived from an analytic representation of a spheroid.

Moreover, when it comes to using SNSPD as part of an optical assembly, 3D-printed lenses can greatly relax the associated alignment accuracy requirements, which is particularly important for cryogenic systems, where mechanical stress during cool-down can lead to significant misalignment. In particular, 3D-printed lenses can help to greatly simplify the coupling of SNSPD to single-mode fibers (SMF), as used in many experiments. To this end, lenses printed both on the SNSPD and on the fiber facets allow to enlarge the diameter of the free-space beam and thus to increase the resilience with respect to translational movements of the components [28,29,35]. Lenses printed to the facets of SMF to facilitate coupling have previously been demonstrated in a series of experiments [28,29,35,43–46]. Regarding multi-channel detectors, 3D-printed lenses further offer the possibility to interface on-chip SNSPD arrays to arrays of optical fibers, which are commercially available with standard pitches of, e.g., 127 µm or 250 µm. This leads to greatly improved detector performance and to simplified assembly processes compared to conventional approaches that rely on mounting of optical fibers by means of dedicated micromachined alignment structures [47,48]. Note that 3D-printed structures can also be used for efficient coupling of light into on-chip waveguides [28,30,49,50], which can then be equipped with SNSPD [5,51–53].

## 6. Summary

We have demonstrated a new approach that exploits 3D-printed micro-lenses to increase the effective collection area of superconducting nanowire single-photon detectors (SNSPD) while keeping the nanowire short, thereby overcoming a fundamental design conflict of such devices. In a proof-of-concept experiment, we show that, for a plane-wave-like illumination at a wavelength of 1550 nm, a lens of 60 µm diameter can provide a 100-fold increase of the effective area of a niobium nitrate (NbN) SNSPD with physical area of $4.5 \times 4.5$ µm$^2$. Since the length of the nanowire can remain small, its maximum achievable count rate is high and its geometrical jitter stays low. In addition, under the constraints of realistically achievable film homogeneity and defect density, SNSPD with small active detection areas offer higher fabrication yield. Our approach enables simplified fabrication of extended SNSPD arrays that feature unprecedented effective fill factors while offering high detection efficiency, high photon count rate (PCR), and high fabrication yield.

**Funding**

This work was supported by the Deutsche Forschungsgemeinschaft (DFG, German Research Foundation) under Germany's Excellence Strategy via the Excellence Cluster 3D Matter Made to Order (EXC-2082/1-390761711) as well as through the DFG project "Physical limits for sensitivity of a monolithic terahertz superconducting sensor based on a galvanically isolated nanobridge" (# 388956995), by the Bundesministerium für Bildung und Forschung (BMBF) via the joint project PRIMA (# 13N14630) and the project DiFeMiS (# 16ES0948), which is part of the programme "Forschungslabore Mikroelektronik Deutschland (ForLab), by the European Research Council (ERC Consolidator Grant 'TeraSHAPE'; # 773248), by the H2020 Photonic Packaging Pilot Line PIXAPP (# 731954), by the Alfried Krupp von Bohlen und Halbach Foundation, and by the Karlsruhe School of Optics and Photonics (KSOP).

**Acknowledgments**

The authors would like to thank S. Doerener, M. Merker, A. Schmid and S. Wuensch for fruitful discussions.

**Disclosures**

P.-I.D. and C.K. are co-founders and shareholders of Vanguard Photonics GmbH and Vanguard Automation GmbH, start-up companies engaged in exploiting 3D nanoprinting in the field of photonic integration and assembly. Y.X., M.B., P.-I.D., and C.K. are co-inventors of patents owned by Karlsruhe Institute of Technology (KIT) in the technical field of the publication. M.B. is now an employee of Nanoscribe GmbH, a company selling 3D lithography systems.

**Data availability**

Data underlying the results presented in this paper may be obtained from the authors upon reasonable request.

**Appendix**

*A. Mathematical models and methods for analysis of spheroidal lenses*

The ideal lens shapes considered in Section 3.1 are obtained by revolving a Cartesian oval curve about its axis of symmetry. This Cartesian oval curve comprises all points having the same linear combination of distances $d_1$ and $n_{\text{lens}} \times d_2$ from two fixed points $F_1$ and $F_2$, respectively, where $n_{\text{lens}}$ denotes the refractive index of the lens and where $F_2$ lies within the lens. The points $F_1$ and $F_2$ are referred to as the foci of the lens, where any ray passing through $F_1$ will be refracted to pass through $F_2$ as well. In the special case considered in Section 3.1, where light is incident as a plane wave, the focus $F_1$ moves to infinity, and the Cartesian oval turns into an ellipse, which, by revolution about the optical axis, defines the associated spheroidal lens. The geometrical focus $S_1$ of this spheroid, see Fig. 2(a), then coincides with the internal focus $F_2$. Starting from this lens shape, we calculate the effective collection area and the minimum achievable spot size.

### A.1 Effective collection radius

In the following, we assume a plane-wave-like illumination with approximately constant intensity over the cross section of the lens. To calculate the effective collection area $A_C$ and the associated effective collection radius $r_C$ of such a spheroid, we consider rays that hit the lens surface at a normalized radial position $\rho = r/a$ under an incidence angle $\alpha_{\text{vac}}(\rho)$ with respect to the local surface normal, see Fig. 2 (a). We denote the Fresnel power transmission at this radial position with $\overline{T}(\alpha_{\text{vac}}(\rho))$, where the overbar denotes the average of the power transmission for p- and s-polarized light. The effective collection area $A_C$ of the lens is defined as the cross-sectional area perpendicular to the optical axis, which would collect the same optical power as the lens itself, and can be calculated by integrating the position-dependent power transmission over the transverse cross section of the spheroidal lens with minor half-axis $a$,

$$A_C = \pi r_C^2 = \int_0^{2\pi}\int_0^a \overline{T}\left(\alpha_{\text{vac}}\left(\frac{r}{a}\right)\right) r \, dr \, d\varphi = 2\pi a^2 \int_0^1 \overline{T}(\alpha_{\text{vac}}(\rho)) \, \rho \, d\rho, \tag{15}$$

$$r_C = a\sqrt{2}\sqrt{\int_0^1 \overline{T}(\alpha_{\text{vac}}(\rho)) \rho \, d\rho}. \tag{16}$$

The incident angle $\alpha_{\text{vac}}(\rho)$ is found from considering the surface-normal direction of the spheroid and can be calculated from the equation of an ellipse to be

$$\alpha_{\text{vac}}(\rho) = \arctan\left(\xi \frac{1}{\sqrt{(1/\rho)^2 - 1}}\right). \tag{17}$$

For the plot in Fig. 2d, we numerically evaluate the integral in Eq. (16). For a refractive index $n_{\text{lens}} = 1.5$, we find $r_C \approx 0.94 \times a$.

### A.2 Estimation of minimum spot size

For estimating the minimum achievable spot size, we calculate the vectorial point spread function according to Richards and Wolf [54,55]. We chose this approach since the maximum involved ray angle $\theta_{\text{max}}$ clearly exceeds the validity range of paraxial approximation, see Eq. (4) in the main text. We first consider a uniform *x*-polarized plane-wave illumination at angular frequency $\omega$ propagating along the *z*-direction and use a positive time dependence, i.e., $\exp(\mathrm{j}(\omega t - k_0 z))$, where $k_0 = \omega/c$ is the vacuum wave number and where $c$ denotes the vacuum speed of light. The complex electrical field vector $\underline{\mathbf{E}}(r,\varphi,z)$ is calculated in a cylindrical coordinate system having its origin at the focus $S_1$ within the lens, see Fig. 2(a). The different components of the vectorial point spread function can be expressed by three integrals $I_0$, , and $I_2$ over the ray angle $\theta$, which contain the so-called real-valued pupil apodization function $P(\theta)$, describing the mapping of the incident field amplitudes from a planar to a spherically converging phase front. Denoting the $n$-th-order Bessel function of the first kind as $\mathrm{J}_n(\cdot)$, the E-field can be written as ([54], Eq. (6.5.9) in [55])

$$\underline{\mathbf{E}}(r,\varphi,z) = \mathrm{j}A\left\{\left[I_0 + \cos(2\varphi)I_2\right]\hat{\mathbf{e}}_x + \sin(2\varphi)I_2\hat{\mathbf{e}}_y + 2\mathrm{j}\cos(\varphi)I_1\hat{\mathbf{e}}_z\right\}, \tag{18}$$

where $A$ is a real-valued amplitude and where the integrals $I_0(r,z)$, $I_1(r,z)$, and $I_2(r,z)$ are calculated as

$$I_0(r,z) = \int_0^{\theta_{\max}} P(\theta)\sin\theta\,(1+\cos\theta)\,\mathrm{J}_0(n_{\mathrm{lens}}k_0 r\sin\theta)\,\mathrm{e}^{-\mathrm{j}n_{\mathrm{lens}}k_0 z\cos\theta}\,\mathrm{d}\theta, \tag{19}$$

$$I_1(r,z) = \int_0^{\theta_{\max}} P(\theta)\sin^2\theta\,\mathrm{J}_1(n_{\mathrm{lens}}k_0 r\sin\theta)\,\mathrm{e}^{-\mathrm{j}n_{\mathrm{lens}}k_0 z\cos\theta}\,\mathrm{d}\theta, \tag{20}$$

$$I_2(r,z) = \int_0^{\theta_{\max}} P(\theta)\sin\theta\,(1-\cos\theta)\,\mathrm{J}_2(n_{\mathrm{lens}}k_0 r\sin\theta)\,\mathrm{e}^{-\mathrm{j}n_{\mathrm{lens}}k_0 z\cos\theta}\,\mathrm{d}\theta. \tag{21}$$

In these relations, the upper integration limit $\theta_{\max}$ corresponds to the maximum ray angle inside the lens and is given by Eq. (4) in the main text. For simplicity, we assume that the SNSPD can be modeled as perfect power detector lying in a $z$-normal plane that is only sensitive to the flux of incoming photons, irrespective of polarization. We express the flux of photons incident onto the SNSPD by the $z$-component of the real part of the complex Poynting vector, which turns out to be independent of the azimuthal coordinate $\varphi$ (Eq. (3.22) in [54]),

$$S_z(r,z) = \frac{1}{2}\Re\left\{\underline{\mathbf{E}}\times\underline{\mathbf{H}}^*\right\}\cdot\hat{\mathbf{e}}_z \propto |I_0|^2 - |I_2|^2. \tag{22}$$

To obtain an expression for the pupil apodization function $P(\theta)$, we first consider the so-called ray projection function $g(\theta)$, which can be expressed by the dependence of the radial position $r(\theta)$ of an incident ray on the ray angle within the lens, see Eq. (6.3.1) in [55],

$$g(\theta) = r(\theta)/f. \tag{23}$$

In this relation, $f$ corresponds to the material-sided focal length of the spheroidal lens, see Fig. 2(a) for an illustration of the various quantities. The apodization function can then be derived from energy conservation considerations, see Eq. (6.3.6) in [55],

$$P(\theta) = \left|\frac{g(\theta)g'(\theta)}{\sin\theta}\right|, \tag{24}$$

where $g'(\theta)$ is the derivative of $g(\theta)$ with respect to $\theta$. We compute the underlying relation between $r$ and $\theta$ from the equation of the spheroidal lens surface and geometrical considerations, see Fig. 2a, and find

$$r(\theta) = \left( \frac{e + \sqrt{1 + \tan^2 \theta}}{1 + \xi^2 \tan^2 \theta} \right) b \tan \theta ,\qquad(25)$$

where $b$ and $e$ denote the major half axis and the eccentricity of the spheroid, respectively. Using $f = b \times (e+1)$, see Eq. (8), leads to

$$g(\theta) = \left( \frac{e + \sqrt{1 + \tan^2 \theta}}{1 + \xi^2 \tan^2 \theta} \right) \frac{\tan \theta}{e+1} .\qquad(26)$$

For each local incidence angle $\alpha_{\text{vac}}$ with respect to the surface normal of the lens, we additionally consider the Fresnel power transmission $\overline{T}(\alpha_{\text{vac}}(r(\theta)/a))$. Note that this represents a simplifying approximation since $\overline{T}(\alpha_{\text{vac}}(r(\theta)/a))$ refers to the average of the power transmission for p- and s-polarized light, whereas our derivation here was based on an incoming *x*-polarized plane-wave. With this simplification, the modified apodization function can be written as

$$P(\theta) = \left| \frac{g(\theta) g'(\theta)}{\sin \theta} \right| \sqrt{\overline{T}(\alpha_{\text{vac}}(r(\theta)/a))} .\qquad(27)$$

The direct relationship between the local incidence angle $\alpha_{\text{vac}}$ and the ray angle $\theta$ is found by geometrical considerations and Snell's law, see Fig. 2a, resulting in an implicit equation, see Eq. (3) in the main text, which we solve numerically. With this, we can finally evaluate the integrals $I_0(r,z)$, $I_1(r,z)$, and $I_2(r,z)$ in Eqs. (19), (20), and (21) and calculate the resulting intensity distribution $S(r, z=0)$ according to Eq. (22). The resulting second-moment-radius $w_0 = 2\sigma$ plotted in Fig. 2b in the main text is obtained from the variance $\sigma^2$ of the intensity distribution along the transverse direction,

$$\sigma^2 = \frac{\int_0^{2\pi}\int_0^\infty r^2 \cos^2(\varphi)\, S(r,z=0)\, r\, dr\, d\varphi}{\int_0^{2\pi}\int_0^\infty S(r,z=0)\, r\, dr\, d\varphi} = \frac{1}{2} \frac{\int_0^\infty r^2 S(r,z=0)\, r\, dr}{\int_0^\infty S(r,z=0)\, r\, dr} .\qquad(28)$$

## B. Intrinsic detection efficiency (IDE) of SNSPD

To draw quantitative conclusions from experimental data, where only the system detection efficiency SDE is accessible, we have to account for the parametric dependency of the intrinsic detection efficiency IDE on the wavelength $\lambda$ and the bias current $I_b$. Due to geometrical inhomogeneity, granularity of the nanowire material, and thermally activated fluctuations of its superconducting state, the nanowire randomly switches to the normal conducting state – even in absence of photons – if the bias is close to a so-called switching current $I_{sw}$, which is usually noticeably lower than the critical current, $I_{sw}/I_c < 0.7$. The bias-current dependency of the intrinsic detection efficiency $\text{IDE}(\lambda, I_b/I_{sw})$ is hence typically described using the relative bias $I_b/I_{sw}$. Regarding the wavelength-dependence for a given bias current, the the intrinsic detection efficiency $\text{IDE}(\lambda, I_b/I_{sw})$ is usually equal to unity up to a bias-dependent cut-off wavelength $\lambda_c(I_b)$. For wavelengths larger than $\lambda_c$, $\text{IDE}(\lambda, I_b/I_{sw})$ decays, where the exact slope of the decay again depends on $I_b$. At operating temperatures above 4 K, it is not always possible to achieve a cut-off wavelength $\lambda_c$ larger than the wavelength $\lambda$ of the test signal by solely increasing the bias current, and the intrinsic detection efficiency remains usually smaller than optimum, $\text{IDE}(\lambda) < 1 \quad \forall \quad I_b < I_{sw}$. The function $\text{IDE}(\lambda, I_b/I_{sw})$ can be experimentally inferred from measured photon count rates PCR at different wavelengths $\lambda$ and bias currents $I_b$, if the device is operated under a constant photon flux and if the optical coupling efficiency OCE and the absorption efficiency ABS can be assumed to be wavelength-independent, see Eq. (1),

$$\text{PCR}(\lambda, I_b/I_{sw}) \propto \text{SDE}(\lambda, I_b/I_{sw}) = \text{OCE} \times \text{ABS} \times \text{IDE}(\lambda, I_b/I_{sw}). \tag{29}$$

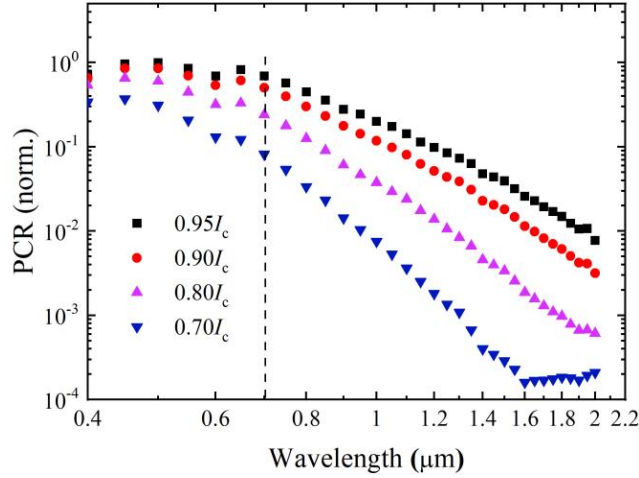

**Fig. 6.** Wavelength- and bias-dependent photon count rates of a NbN detector similar to the ones used in our experiment. The device is operated under constant photon flux, and the count rates are normalized to the respective maximum in the limit of low wavelengths. Assuming constant optical coupling efficiency OCE and constant absorption efficiency ABS, these curves represent the dependence of the intrinsic detection efficiency (IDE) on wavelength and bias current. Beyond the bias-current-dependent cutoff wavelength $\lambda_c$ the curves decay and roughly follow straight lines in the semi-logarithmic plot, with characteristic slopes that monotonically decreases in magnitude with increasing relative bias $I_b/I_{sw}$. At a temperature of 4.2 K and bias currents $I_b = 0.95 I_{sw}$, the cut-off wavelength amounts to $\lambda_c \approx 700$ nm [27], indicated by a dashed vertical line.

Figure 6 depicts such a measurement of the photon count rate for an NbN detector that is similar to the devices used in our experiment. The photon count rates are normalized to the maximum values found in the low-wavelength limit, corresponding to an intrinsic detection efficiency $\text{IDE}(\lambda) \approx 1$ below the respective cut-off wavelength $\lambda_c(I_b)$. At a temperature of 4.2 K and a relative bias currents $I_b/I_{sw} = 0.95$, the cut-off wavelength amounts to $\lambda_c \approx 700$ nm [27], indicated by a dashed vertical line in Fig. 6. Beyond the cutoff wavelength $\lambda_c$, the IDE roughly follows a straight line in the semi-logarithmic plot, with the magnitude of the slope monotonically decreasing with increasing relative bias $I_b/I_{sw}$. To adjust bias currents for equal IDE for lensed and reference detector, $\text{IDE}_{lens} = \text{IDE}_{ref}$, we use equal relative bias currents $I_b/I_{sw}$, see Section 5.2. To further support the notion that this indeed leads to equal IDE, we perform separate reference measurements using backside illumination of the lensed (lens) detector and the reference (ref) detector at two alternating wavelengths $\lambda_1 = 850$ nm and $\lambda_2 = 1550$ nm, and experimentally confirm that the ratios of photon count rates at the two wavelengths are identical,

$$\frac{\text{PCR}_{lens}(\lambda_1)}{\text{PCR}_{ref}(\lambda_1)} = \frac{\text{PCR}_{lens}(\lambda_2)}{\text{PCR}_{ref}(\lambda_2)}. \tag{30}$$

If those two ratios are identical, we can infer equal intrinsic detection efficiency even if the backside illumination through the multimode fiber might not be sufficiently homogeneous such that the rate of received photons $R_{\text{r,lens}}$ at the backside of the lensed and $R_{\text{r,ref}}$ at the backside of the reference detector differ. This can be shown by assuming equal and wavelength-independent absorption efficiency ABS, and by substituting $\text{PCR}_{\text{lens}}(\lambda_m) = R_{\text{r,lens}}(\lambda_m) \times \text{ABS} \times \text{IDE}_{\text{lens}}(\lambda_m)$, see Eq. (1), and analogously $\text{PCR}_{\text{ref}}(\lambda_m) = R_{\text{r,ref}}(\lambda_m) \times \text{ABS} \times \text{IDE}_{\text{ref}}(\lambda_m)$. Equation (30) can then be written as

$$\frac{R_{\text{r,lens}}(\lambda_1)}{R_{\text{r,ref}}(\lambda_1)} \frac{\text{IDE}_{\text{lens}}(\lambda_1)}{\text{IDE}_{\text{ref}}(\lambda_1)} = \frac{R_{\text{r,lens}}(\lambda_2)}{R_{\text{r,ref}}(\lambda_2)} \frac{\text{IDE}_{\text{lens}}(\lambda_2)}{\text{IDE}_{\text{ref}}(\lambda_2)}. \tag{31}$$

Assuming at least a wavelength-independent ratio of received photons, $R_{\text{r,lens}}(\lambda_1)/R_{\text{r,ref}}(\lambda_1) = R_{\text{r,lens}}(\lambda_2)/R_{\text{r,ref}}(\lambda_2)$, we can further simplify and rearrange Eq. (31) to find

$$\frac{\text{IDE}_{\text{lens}}(\lambda_1)}{\text{IDE}_{\text{lens}}(\lambda_2)} = \frac{\text{IDE}_{\text{ref}}(\lambda_1)}{\text{IDE}_{\text{ref}}(\lambda_2)}. \tag{32}$$

Equal ratios in Eq. (30) thus imply equal decay slopes in the semi-logarithmic plot of Fig. 6. We can hence conclude that the lensed and reference detector are operated on the same characteristic curve and thus have the same intrinsic detection efficiency IDE if Eq. (30) is fulfilled, i.e., if the ratios of photon count rates at the two wavelengths are identical.